\documentstyle[aps,amssymb]{revtex}

\begin{document}
\title{A novel quantum key distribution scheme with orthogonal product states }
\author{Guo-Ping Guo, Chuan-Feng Li\thanks{%
Electronic address: cfli@ustc.edu.cn }, Bao-Sen Shi, Jian Li, and Guang-Can
Guo\thanks{%
Electronic address: gcguo@ustc.edu.cn }}
\address{Laboratory of Quantum Communication and Quantum Computation and Department\\
of Physics, University of Science and Technology of China, Hefei 230026,\\
People's Republic of China}
\maketitle

\begin{abstract}
\baselineskip12pt{}The general conditions for the orthogonal product states
of the multi-state systems to be used in quantum key distribution (QKD) are
proposed, and a novel QKD scheme with orthogonal product states in the $%
3\times 3$ Hilbert space is presented. We show that this protocol has many
distinct features such as great capacity, high efficiency. The
generalization to $n\times n$ systems is also discussed and a fancy
limitation for the eavesdropper's success probability is reached.

PACS number(s): 03.67.Dd, 03.65.Bz
\end{abstract}

\section{Introduction}

\baselineskip12pt{}Cryptography is created to satisfy the people's desire of
transmitting secret messages. With the development of the quantum
computation, especially the proposal of Shor's algorithm\cite{f1}, the base
of the most important classic cryptographic scheme was shocked. But at the
same time, the principles of quantum mechanics have also shed new light on
the field of cryptography as these fundamental laws guarantee the secrecy of
quantum cryptosystems. Any intervention of an eavesdropper, Eve, must leave
some trace which can be detected by the legal users of the communication
channel. All kinds of quantum key distribution (QKD) schemes, such as BB84
protocol\cite{f2}, B92 protocol\cite{f3}, and the EPR scheme\cite{f4} have
been proposed. Recently, quantum cryptography with $3$-state systems was
also introduced\cite{f5}. Experimental research on QKD is also progressing
fast, for instance, the optical-fiber experiment of BB84 and B92 protocols
have been realized up to 48 km\cite{f6}, and QKD in free space for B92
scheme has been achieved over 1 km distance\cite{f7}.

In paper\cite{f8}, Lior and Lev presented a quantum cryptography based on
orthogonal states firstly. Then there is quantum cryptographic scheme
involving truly two orthogonal states \cite{f9}. The basic technic is to
split the transfer of one-bit information into two steps, ensuring that only
a fraction of the bit information is transmitted at a time. Then the
non-cloning theorem of orthogonal states\cite{f10} guarantee its security.
Based on the impossibility of cloning nonorthogonal mixed states, the
no-cloning theorem of orthogonal states says that the two (or more)
orthogonal states $\rho _i\left( AB\right) $ of the system composed of $A$
and $B$ cannot be cloned if the reduced density matrices of the subsystem
which is available first (say $A$) $\rho _i\left( A\right) =Tr_B\left[ \rho
_i\left( AB\right) \right] $ are nonorthogonal and nonidentical, and if the
reduced density matrices of the second subsystem are nonorthogonal. It is a
very surprising result since it means that entanglement is not vital for
preventing cloning of orthogonal states. In the case of a composite system
made of two subsystems, if the subsystem are only available one after the
other, then there are various cases that orthogonal states cannot be cloned.

For the multi-state systems, Bennett {\it et al.} have shown that there are
orthogonal product pure states in the $3\times 3$ Hilbert space and proved
that these states may have some degree of nonlocality without entanglement%
\cite{f11}. There was also experimental demonstration of three mutually
orthogonal polarization states\cite{f12}, where biphotons are used as
multi-state systems.

We propose the general conditions for the orthogonal product states of the
multi-state composite systems to be used in QKD, then present a QKD scheme
with the orthogonal product states of $3\times 3$ system which has several
distinct features, such as high efficiency and great capacity. The
generalization to the $n$-state systems, and eavesdropping is analyzed where
a peculiar limitation, $1/2,$ for the success probability of an efficient
eavesdropping strategy is found as{\it \ }$n${\it \ }becomes large enough.

\section{The QKD Scheme with Orthogonal Product States}

In the present QKD scheme with orthogonal product states in the $n\times n$
Hilbert space, the transmission processing is same as the QKD scheme with
common orthogonal states\cite{f8}. The information is encoded in the
holistic state of the two particles, and these two particles are sent
separately to ensure that any eavesdropper can not hold both particles at
the same time. Since only orthogonal product state are employed, operations
on one subsystem have no effect to the other. There are some basic
conditions for any set of orthogonal product states in the $n$-state
composite systems to be used in the present QKD scheme: for any density
matrix of any subsystem, $\rho _i\left( P\right) ,$ there must be at least
one $\rho _j\left( P\right) $ which is both nonidentical and nonorthogonal
to $\rho _i\left( P\right) .$ ($P$ represents subsystem $A$ or $B$. $i$ and $%
j$ represent different states of the set.) Then from the point of any
subsystem\cite{f10}, the standard non-cloning theorem \cite{f13} is
satisfied, this guarantees the security of the protocol. What's more, we can
transmit $2\log _2n$ bits information, double the value of existing QKD
protocol with usual orthogonal states\cite{f8,f9}. It is evident that this
is the maximal information that can be transmitted by the $n\times n$ system.

For $2\times 2$ system, there is obviously no such orthogonal product states
that satisfy the orthogonal states cryptography conditions. The reason is
that if $\rho _0\left( P\right) $ are nonidentical and nonorthogonal to $%
\rho _1\left( P\right) $, then $\rho _0\left( A\right) \otimes \rho _0\left(
B\right) $ can not be orthogonal to $\rho _1\left( A\right) \otimes \rho
_1\left( B\right) .$

Next, we consider the $3\times 3$ system. A general set of orthogonal
product states in this Hilbert space is as following 
\begin{equation}
\left. 
\begin{array}{l}
\Psi _1=\left| 1\right\rangle _A\left( a\left| 1\right\rangle _B+b\left|
0\right\rangle _B\right) , \\ 
\Psi _2=\left| 1\right\rangle _A\left( b^{*}\left| 1\right\rangle
_B-a^{*}\left| 0\right\rangle _B\right) , \\ 
\Psi _3=\left( c\left| 1\right\rangle _A+d\left| 0\right\rangle _A\right)
\left| 2\right\rangle _B, \\ 
\Psi _4=\left( d^{*}\left| 1\right\rangle _A-c^{*}\left| 0\right\rangle
_A\right) \left| 2\right\rangle _B, \\ 
\Psi _5=\left| 2\right\rangle _A\left( e\left| 0\right\rangle _B+f\left|
2\right\rangle _B\right) , \\ 
\Psi _6=\left| 2\right\rangle _A\left( f^{*}\left| 0\right\rangle
_B-e^{*}\left| 2\right\rangle _B\right) , \\ 
\Psi _7=\left( g\left| 0\right\rangle _A+h\left| 2\right\rangle _A\right)
\left| 1\right\rangle _B, \\ 
\Psi _8=\left( h^{*}\left| 0\right\rangle _A-g^{*}\left| 2\right\rangle
_A\right) \left| 1\right\rangle _B, \\ 
\Psi _9=\left| 0\right\rangle _A\left| 0\right\rangle _B,
\end{array}
\right. 
\end{equation}
where $a,$ $b,$ $c,$ $d,$ $e,$ $f,$ $g,$ $h$ are complex number and $\left|
a\right| ^2+\left| b\right| ^2=\left| c\right| ^2+\left| d\right| ^2=\left|
e\right| ^2+\left| f\right| ^2=\left| g\right| ^2+\left| h\right| ^2=1.$

This set of states has been proved to have some degree of nonlocality
without entanglement, when $a=b=c=d=e=f=g=h=1/\sqrt{2}$\cite{f11}. For the
general case, no satisfying proof for the existing of the nonlocality has
been found yet. But if they satisfy the conditions mentioned above, they can
still be used in this QKD scheme.

The process of this QKD scheme is as follows: 1. Alice prepares two
particles $A$ and $B$ randomly in one of the nine orthogonal product states
shown above and sends particle $A$ to Bob, when Bob receives it, he informs
Alice through open classical channel. Then Alice sends out particle $B$.
When particle $A$ and $B$ are both in the hand of Bob, he makes a collective
orthogonal measurement under the basis of Eqs. (1) to determine which state
the two-particle system has been prepared. After a sequence of this
procedure, they can share a random bit string, which is the raw keys. 2. In
order to find possible eavesdropping, Alice and Bob randomly compare some
bits to verify whether the correlations have been destroyed. If it is true
with as high probability as they require, it can be believed that there is
no eavesdropper and all of the rest results can be used as cryptographic
keys. Otherwise, all the keys are discarded and they must be redistributed.

What is vital to this scheme is that Alice sends the second particle only
when the first one reaches Bob to eliminate the possibility of any
eavesdropper to possess the two particles at the same time. This protocol
has some distinct features. As all raw keys except a small portion chosen
for checking eavesdroppers is usable, it is very efficient (nearly $100\%$),
and it has large capacity since $\log _29$ bits information is transmitted
by a $3\times 3$ system.

\section{Eavesdropping and the generalization to the $n$-state system}

We first consider one efficient eavesdropping strategy. In this strategy,
Eve measures the first particle from Alice and sends it to Bob. She measures
the second particle corresponding to the measurement result of the first one
and sends it to Bob. It has been found that with this strategy, Eve has the
success eavesdropping probability as high as $0.94$ for QKD with orthogonal
states in the $2\times 2$ Hilbert space\cite{f14}.

The eavesdropping is as following, Eve intercepts particle $A$ and makes an
orthogonal measurement in the basis $\left\{ \left| 0\right\rangle ,\left|
1\right\rangle ,\left| 2\right\rangle \right\} $. Suppose particle $A$ is
found in state $\left| 1\right\rangle _A$, Eve knows that the two-particle
states of $A$ and $B$ is $\Psi _1$, $\Psi _2$ with probability $1/9$
respectively, or $\Psi _3$, $\Psi _4$ with probability $\left| c\right| ^2/9$
and $\left| d\right| ^2/9,$ respectively. Then she sends it to Bob. When
particle $B$ comes, she intercepts it too and measures it in the basis $%
\left\{ \left| 2\right\rangle _B,\text{ }a\left| 1\right\rangle _B+b\left|
0\right\rangle _B,\text{ }b\left| 1\right\rangle _B-a\left| 0\right\rangle
_B\right\} $ and then sends it to Bob. If Eve sees that particle $B$ is in $%
a\left| 1\right\rangle _B+b\left| 0\right\rangle _B$ or $b\left|
1\right\rangle _B-a\left| 0\right\rangle _B$, then obviously she knows that
the two-particle state is $\Psi _1$or $\Psi _2$. In this case she is lucky
enough to conceal from Alice and Bob for the two-particle state is not
disturbed. And if Eve finds particle $B$ in state $\left| 2\right\rangle _B$%
, then the two-particle state is $\Psi _3$ or $\Psi _4$, which has collapsed
to $\left| 1\right\rangle _A\left| 2\right\rangle _B$, then Bob has the
partial probability of $\left| c\right| ^2$ or $\left| d\right| ^2$ to find
the two-particle state in $\Psi _3$ or $\Psi _4,$ respectively. So it is
clear that for the case Eve measures the particle $A$ in the state $\left|
1\right\rangle _A$, the probability for he to eavesdrop without being
detected is $\left( 2+\left| c\right| ^4+\left| d\right| ^4\right) /9.$
Analyzed in the same way, the total probability that Eve eavesdrops the key
information without being detected is 
\begin{equation}
P_3=5/9+2(\left| c\right| ^4+\left| d\right| ^4+\left| h\right| ^4+\left|
g\right| ^4)/9.
\end{equation}
Then it is evident that $P_3$ gets the minimal value of $7/9$ when $\left|
c\right| ^2=\left| d\right| ^2=\left| h\right| ^2=\left| g\right| ^2=1/2$.

We depict this set of states in the $3\times 3$ Hilbert space in a visual
graphical way as Figure 1 shows. The four dominoes represent the four pairs
of states that involve superposition of the basis states $\left|
0\right\rangle $, $\left| 1\right\rangle $ or $\left| 2\right\rangle $. It
is obvious that this figure is $4$-fold rotation symmetric, and we will show
later that this symmetry is one of the basic requirements for there may be
two symmetrical eavesdropping strategies. From this figure we can obviously
see that all the states included in one row, where particle $A$ is in basis
states, can be eavesdropped without being detected in this strategy. But for
all other states, there is only certain probability, the quartic of
probability amplitude of the superposition states of particle $A$ in the
basis states, for Eve's success eavesdropping. Then the total probability is 
$P_3$. Now let's consider the case of $n\times n$ system. Since we only
utilize orthogonal product states, and any superposition of $n$ basis states
that covers all the grids of any row (any column) in the graphic depiction
will surely be distinguished by Eve which will have no use in the present
QKD scheme. In other words, any set of states including such superposition
state will violate the above conditions, then the possible superposition
states can just be the ones of less than $n$ basis states which cover no
more than $n-1$ grids of the figures. The $4\times 4$ system can be depicted
in Figure 2. We can see that the only difference between the figures of the $%
3\times 3$ and $4\times 4$ systems is that there are four states in the
center of the $4\times 4$ system, which can be eavesdropped without being
detected. Then generalized straightforwardly, the $2n\times 2n$ system can
be analyzed in the similar way as the $(2n-1)\times (2n-1)$ system. Thus
here we take the $(2n-1)\times (2n-1)$ system for example. $(n=2,3,4,...)$.

The figure 3 depicts a set of orthogonal product states of the $5\times 5$
system, which is generalized straightforwardly from $3\times 3$ system. For
any complete set of orthogonal product states, the success probability for
Eve to eavesdropping without being detected is 
\begin{equation}
P_5=\left[ 9P_3+8+(\sum_{k,i,J}\left| \alpha _{J_{i_k}\left| 3\right\rangle
_B}\right| ^4+\sum_{k,i,J}\left| \alpha _{J_{i_k}\left| 4\right\rangle
_B}\right| ^4)\right] /25,
\end{equation}
where $\alpha _{J_{i_k}\left| l\right\rangle _B}$ is the probability
amplitude of the superposition states of particle $A$ in the basis states
and $\sum\limits_J\alpha _{J_{i_k}\left| l\right\rangle _B}^2=1$ for any $%
\left| l\right\rangle _B,$ $k$ and $i$, and $\left| l\right\rangle _B$ means
that particle $B$ is in the basis state $\left| l\right\rangle _B$, $k$
denotes dominoes in line $\left| l\right\rangle _B$, $i_k$ denotes the
superposition states in domino $k$, and $J_{i_k}$ denotes the basis states
that superposition state $i_k$ involves, $P_3$ is Eve's success probability
for the $3\times 3$ system and $8/25$ is the probability for states in row $%
\left| 3\right\rangle _A$ and $\left| 4\right\rangle _A$. We know from the
graph that the dominoes in column $\left| 3\right\rangle _B$ and $\left|
4\right\rangle _B$ cover $4$ grids in total (If $5$ grids are covered, all
the states in this column can be eavesdropped without being detected.), and
we have shown that $J_{i_k}\leq 4$. It can be proved easily that $P_5$
reaches its minimal value $17/25$ when $P_3$ gets its minimal value and only
one domino is in column $\left| 3\right\rangle _B$ or $\left| 4\right\rangle
_B$ and $\left| \alpha _{J_{i_k}\left| l\right\rangle _B}\right| ^2=1/4$.
This corresponds to the set of states depicted in Figure 3. Then for the $%
(2n+1)\times (2n+1)$ system, the success probability is 
\begin{equation}
P_{2n+1}=\left[ (2n-1)^2\times P_{2n-1}+4n+(\sum_{k,i,J}\left| \alpha
_{J_{i_k}\left| 2n-1\right\rangle _B}\right| ^4+\sum_{k,i,J}\left| \alpha
_{J_{i_k}\left| 2n\right\rangle _B}\right| ^4)\right] /(2n+1)^2.
\end{equation}
The minimal probability can be obtained similarly: 
\begin{equation}
\min \{P_{2n+1}\}=\left[ (2n-1)^2\times \min \{P_{2n-1}\}+4n+2\right]
/(2n+1)^2=1/2+(1+4n)/2(2n+1)^2,
\end{equation}
where $n\geqslant 1$. We can deduce immediately that this value approaches $%
1/2$ when $n$ gets large enough.

Of course, there are other ways of plotting the graph symmetrically. But
this set of states is of the most secure. In fact, the value of $\min
\{P_{2n+1}\}$ can be deduced straightforwardly from the symmetry of the
plot. For any vertical domino contributes $1/(2n+1)^2$ to this value, any
state in the horizontal dominoes and the state in the center each
contributes $1/(2n+1)^2$. Due to the symmetry, there are at least $2n$
vertical or horizontal dominoes and $\left[ (2n+1)^2-1\right] /2$ states in
the horizontal dominoes.

For the $2n\times 2n$ system, the same result can be reached. That is to say
there is a limitation in the probability of the success eavesdropping when
the Hilbert space becomes large enough. And it is evident that in this
strategy only particle $A$ may be demolished, and particle $B$ is not
infected at all. The function of the operation to $B$ which is depended on
the result of the measurement on $A$ is just to extract more information.

Eve may adopt the complementary eavesdropping strategy, in which Eve try to
eavesdrop some information by intercepting and operating only on the second
particle $B$, which may cause demolition to it. Then for the set of states
in $n\times n$ systems, whose graphic depictions are $4$-fold rotation
symmetric, the probability to eavesdrop some information without being
detected is equal to that of the first strategy, i.e., $P_n$. But for those
states without such symmetry, it can be verified that one of the success
probabilities for the complementary strategies is larger than $P_n$. So we
employ the symmetric states in the present scheme.

Of course, there are other strategies, for example, she can hold up the
first particle $A$ and send out a substitute particle $C$ to Bob. When $B$
comes, she makes a collective measurement under the two-particle orthogonal
basis, then sends out a particle $D$ in the state of $B$. In this strategy,
Eve can eavesdrop the information entirely, but the probability for she to
pass the checking process is only $1/n$, which tends to zero, for the state
of particle $C$ is randomly chosen from a $n$ Hilbert space.

\section{Conclusion}

We have proposed the general conditions for the orthogonal product states to
be used in QKD, then presented a QKD scheme with the orthogonal product
states of $3\times 3$ system which has several distinct features, such as
high efficiency and great capacity. The generalization to the $n$-state
systems, and eavesdropping is analyzed where a peculiar limitation, $1/2,$
for the success probability of an efficient eavesdropping strategy is found
as{\it \ }$n${\it \ }becomes large enough.

{\bf Acknowledgment} This work was supported by the National Natural Science
Foundation of China.

{\bf Figure Captions:}

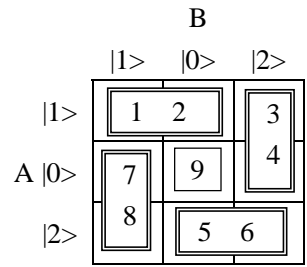
Figure 1: The graphical depiction of the set of orthogonal product states in
the $3\times 3$ Hilbert space.

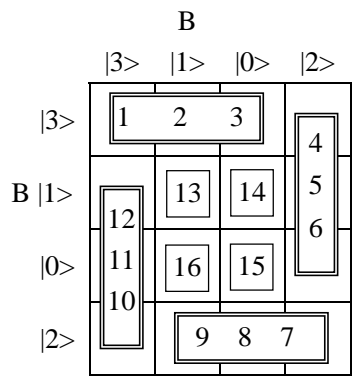
Figure 2: The graphical depiction of the set of orthogonal product states in
the $4\times 4$ Hilbert space.

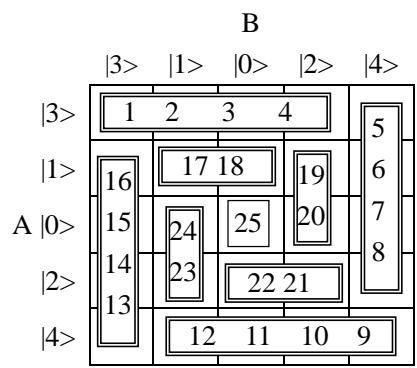
Figure 3: The graphical depiction of the set of orthogonal product states in
the $5\times 5$ Hilbert space.

\end{document}